\documentclass[pra,twocolumn,showpacs,nofootinbib,floatfix,superscriptaddress]{revtex4}

\usepackage{amsmath,amsfonts,amssymb,bm}
\usepackage{dcolumn}
\usepackage[final]{graphicx}
\usepackage{bm,tabularx}
\begin{document}

\title{Parametric squeezing amplification of Bose-Einstein condensates}

\author{Georg J\"ager}
\affiliation{Institut f\"ur Physik, Karl-Franzens-Universit\"at Graz, Universit\"atsplatz 5, 8010 Graz, Austria}

\author{Tarik Berrada}
\author{J\"org Schmiedmayer}
\author{Thorsten Schumm}
\affiliation{Vienna Center for Quantum Science and Technology, Atominstitut, TU Wien, Stadionallee 2, 1020 Vienna, Austria}

\author{Ulrich Hohenester}
\affiliation{Institut f\"ur Physik, Karl-Franzens-Universit\"at Graz, Universit\"atsplatz 5, 8010 Graz, Austria}

\begin{abstract}
We theoretically investigate the creation of squeezed states of a Bose-Einstein Condensate (BEC) trapped in a magnetic double well potential.  The number or phase squeezed states are created by modulating the tunnel coupling between the two wells periodically with twice the Josephson frequency, i.e., through parametric amplification. Simulations are performed with the multi configurational Hartree method for bosons (MCTDHB).  We employ optimal control theory to bring the condensate to a complete halt at a final time, thus creating a highly squeezed state (squeezing factor of 0.12, $\xi_S^2=-18$ dB) suitable for atom interferometry.  
\end{abstract}

\pacs{03.75.-b,39.20.+q,39.25.+k,02.60.Pn}


\maketitle

\section{Introduction}

In atom chips, Bose-Einstein condensates (BECs) and ultracold atoms become trapped in the vicinity of a solid-state chip~\cite{reichel:11}.  By changing the currents running through the wires mounted on the chip or modifying the strength of additional radio-frequency (rf) fields \cite{folman:02,hofferberth:06}, one can manipulate \cite{krueger:03,schumm:05,hofferberth:06} and measure single quantum systems with extremely high precision.  Possible applications range from atom interferometry \cite{haensel:01,andersson:02,wang:05,schumm:05,integrated}, over quantum gates \cite{calarco:00,charron:06,treutlein:06} and resonant condensate transport \cite{paul:05}, to nonlinear atom optics \cite{deng:99,orzel:01,campbell:06,perrin:07,twin}. 

In particular atom interferometry has attracted a lot of interest since atoms are massive objects sensitive to gravity.  This opens new ways for measuring the gravitational constant~\cite{rosi2014}, detection of gravitational waves, or the search for dark energy~\cite{hamilton2015}.  Using non-classical (squeezed) states brings the measurement sensitivity below the quantum noise limit~\cite{cronin:09}.  Squeezed atom number states are typically created through condensate splitting and manipulation of the condensate around the point where the tunnel coupling strength becomes comparable with the nonlinear atom-atom interaction~\cite{riedel:10,ober2,ober1}.  Possible routes towards squeezing are based on quasi-adiabatic splitting~\cite{javanainen:99} or one-axis twisting~\cite{kitagawa:93}.

It is often advantageous to seek for fast squeezing, for instance to achieve measurement series with high repetition rates or to suppress dephasing losses due to thermally excited atoms.  In \cite{grond09,julian09} we demonstrated fast squeezing protocols that were obtained by using optimal control theory (OCT)~\cite{oct1,glaser:15}, a mathematical device allowing for optimization of certain control objectives.  OCT protocols were succesfully implemented in atom chip experiments for twin-atom production~\cite{twin} and interferometry~\cite{nonclassical}.

In this paper we theoretically investigate the generation of squeezed states in a split BEC through parametric amplification.  For a harmonic oscillator, parametric amplification can be achieved by modulating the spring constant with twice the resonance frequency, leading to an exponential increase of the oscillator's amplitude~\cite{landau}.  Similarly, modulating the tunnel coupling strength with twice the Josephson frequency leads to an exponential increase of number and phase fluctuations.  To achieve fast squeezing, say on a time scale of 10 ms, one needs rather large tunnel coupling modulations which lead to additonal wavefunction oscillations of the split condensate, thus rendering the state useless  for further interferometry once the wells become separated.  We demonstrate that a final splitting stage of 2 ms, optimized with OCT, brings the condensate at a final time to halt and freezes the system in a highly squeezed state.

The motivation of this work lies in a direct experimental implementation.  While the combined parametric amplification and splitting scheme investigated in this work leads to a slightly better squeezing compared to previous work~\cite{kitagawa:93,grond09,julian09}, it is additionally simpler to implement and facilitates state tomography by releasing the condensate at different times and recording the time-of-flight images~\cite{twin,inversion}.  

We have organized our paper as follows.  In Sec.~\ref{sec:twomode} we discuss BEC interferometry and squeezing within a two-mode model and introduce a convenient Bloch sphere visualization for the many-body wavefunction.  Squeezing through parametric amplification is discussed in Sec.~\ref{sec:amplification} within the framework of the multi configurational Hartree method for bosons (MCTDHB)~\cite{mctdhb}, which allows for the consideration of both wavefunction and atom number dynamics.  We identify the pertinent parameters that lead to fast and efficient squeezing amplifications.  In Sec.~\ref{sec:trap} we employ the OCT framework to derive control ramps that freeze the condensate in a state with high number squeezing.  Finally, Sec.~\ref{sec:summary} provides a short summary.

\section{Two-mode model}\label{sec:twomode}

\subsection{BECs in double wells}\label{sFec:becindw}

\begin{figure}
\includegraphics[width=0.7\columnwidth]{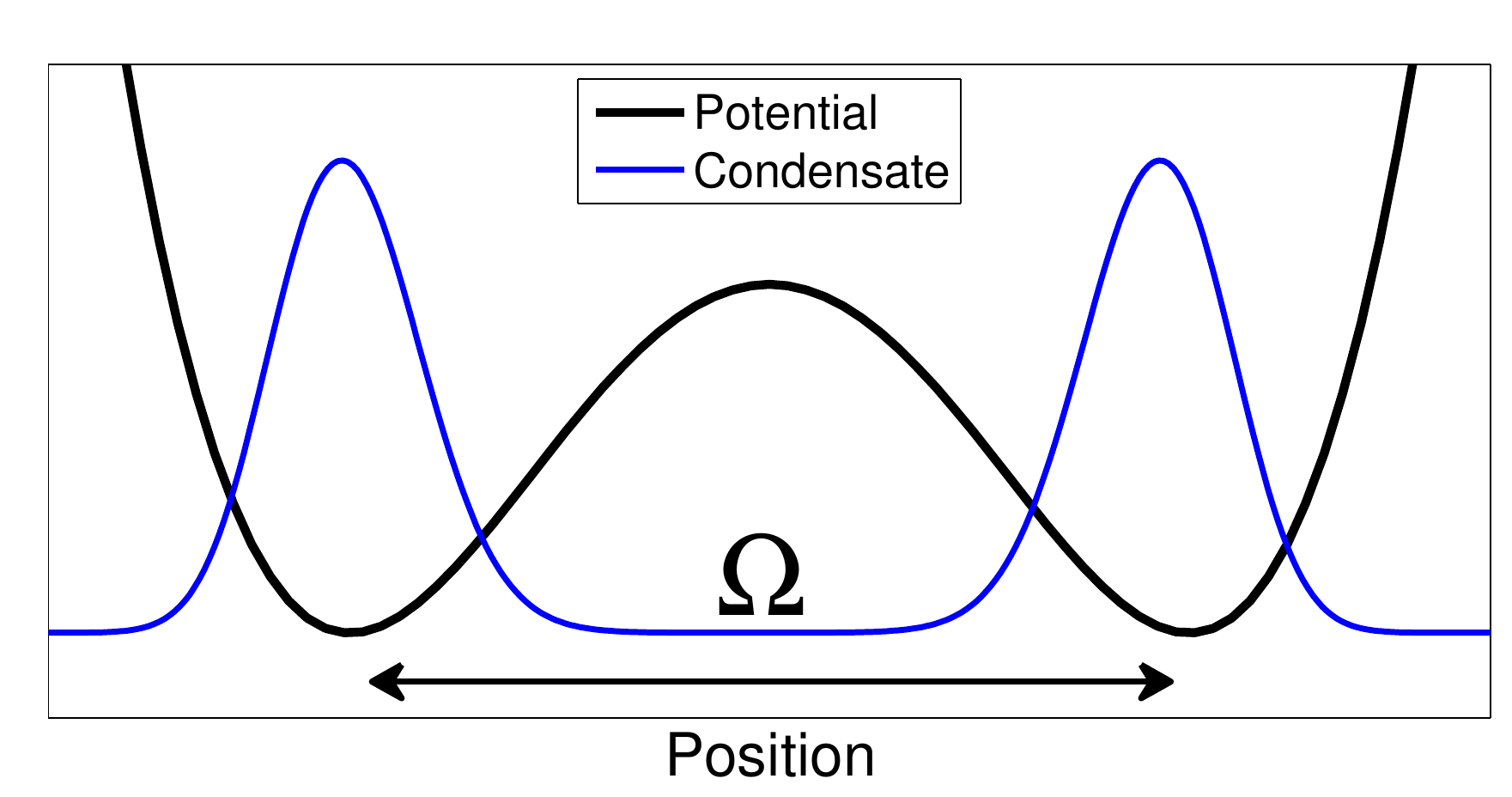}
\caption{(Color online) Schematic of a BEC wave function in a double well potential.  Transforming a single well slowly into a double well produces a split BEC that can be used for interferometry.  In the two-mode model atoms can reside in either the left or right well.  Tunneling promotes atoms between the two wells, and the interwell distance allows control over the tunneling strength $\Omega$. }
\label{dwintro}
\end{figure}

For the purpose of interferometry, we consider a 1d representation of a BEC in a double well trap, as depicted in Fig.~\ref{dwintro}.  We assume that the trap is produced by the magnetic fields generated by an atom chip~\cite{reichel:11}, which allow us to transform the potential from a single to a double well, thus creating a split BEC, and to change the distance between the two wells~\cite{lesanovsky} in order to control the interwell coupling.  Within the field of BEC interferometry, in the waiting phase the atoms in the two wells are decoupled and acquire different phases due to interactions with some external (classical) probe, such as gravity or magnetic fields.  The phase shifts are finally read out through BEC interference.

The physics of double-well BECs is conveniently described in terms of a two-mode model, similar to Josephson junctions \cite{jojunc}, where each atom can either reside in the left or the right well.  With $N$ atoms in the BEC, we can map the model to a spin $N/2$ system that captures many phenomena of double-well BECs.  We introduce the pseudo spin operators~\cite{pseudospin}
\begin{align}
{J_x}=\frac{1}{2} \left( a_l^\dagger a_r + a_r^\dagger a_l \right)\phantom{\,,}\\
{J_y}=\frac{i}{2} \left( a_l^\dagger a_r - a_r^\dagger a_l \right)\phantom{\,,}\\
{J_z}=\frac{1}{2} \left( a_l^\dagger a_l - a_r^\dagger a_r \right)\,,
\end{align}
with $a_{l/r}$ and $a_{l/r}^\dagger$ being the annihilation and creation operators for an atom in the left/right well, respectively.  These operators have the following physical interpretations:  ${J_x}$ exchanges an atom between the left and the right well, and ${J_y}$ and $J_z$ measure the phase difference and atom number imbalance between the two wells, respectively.  With these operators we can write down a model  Hamiltonian in the form \cite{hamil1,hamil2}
\begin{equation}
 H = -\Omega\, J_x + 2\kappa\, J_z^2 \,,
\label{ham}
\end{equation}
where $\Omega$ is the tunneling energy, accounting for the interwell tunneling, and $\kappa$ is the charging energy describing the nonlinear interaction between atoms.  For the interwell distances of our present concern, $\Omega$ can be assumed to be approximately proportional to the distance of the two wells (see Fig.~\ref{dwintro}), while $\kappa$ has in general a more complicated behavior.  Both quantities can be computed within the Gross-Pitaevskii framework~\cite{smerzi99}.

\begin{figure*}
\includegraphics[width=1.8\columnwidth]{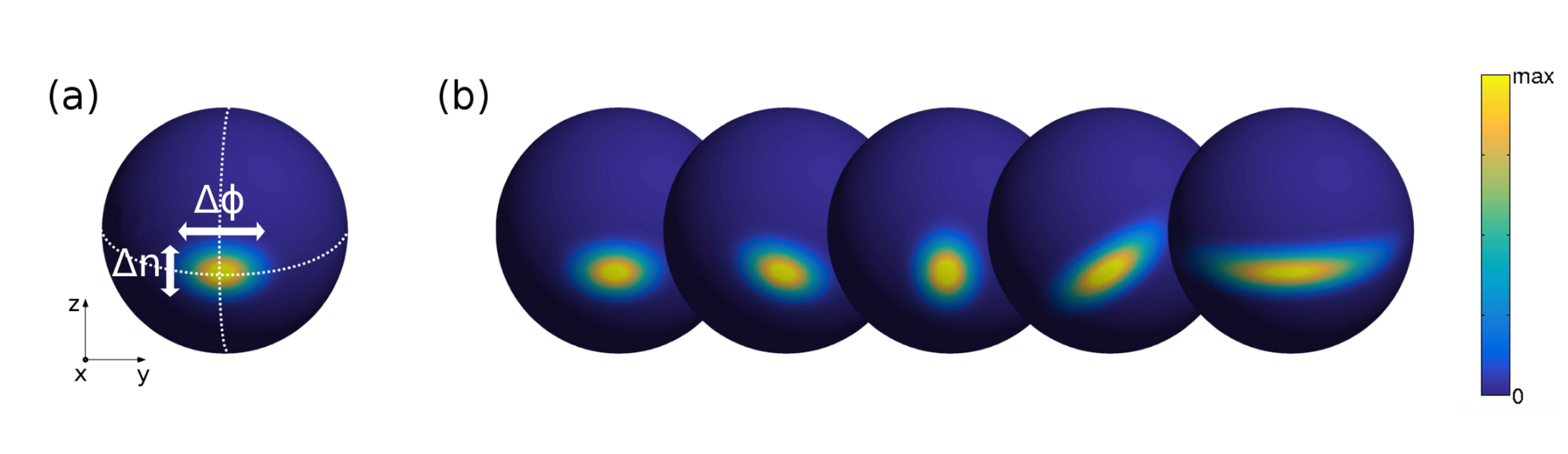}
\caption{(Color online) (a) State of a double-well BEC depicted on the Bloch sphere.  For an equal number of atoms in the two wells, the distribution is centered around the equator, the height of the distribution corresponding to number fluctuations and the width to phase fluctuations.  (b) Schematic view of parametric amplification on the Bloch sphere.  During amplification the distribution rotates around the $x$-axis (driven by modulations of the tunneling strength $\Omega$) and becomes more and more elongated under the influence of the atom-atom nonlinearity.  Parametric amplification leads to an alternation between number and phase squeezed states, and the overall squeezing increases over time. }
\label{bsintro}
\end{figure*}

States of a two-level quantum mechanical system (qubit) are conveniently depicted on the Bloch sphere~\cite{bloch,blochsphere}.  Such visualization is also possible for the two-mode model with a rather intuitive interpretation:  A state where all the atoms are in the left or right well corresponds to a Bloch state on the north or south pole.  We introduce $n=(n_l-n_r)/2$ for the atom number imbalance, with $n_{l/r}$ being the number of atoms in the left/right well.  States where the atom number is exactly balanced, $n=0$, are on the equator, and the angle $\varphi$ describes the relative phase between the two wells (see Fig.~\ref{bsintro}).  In addition to the mean values, also the atom number and phase uncertainties $\Delta n$ and $\Delta\phi$ can be seen on the Bloch sphere: $\Delta n$ corresponds to the height and $\Delta\phi$ to the width of the distribution, as shown in Fig. \ref{bsintro}(a).  

For any interferometry experiment, the important observable to be measured in the end is either the relative phase or number imbalance between the wells.  Both measurements are subject to (shot) noise limiting the measurement precision, and thus render states with large $\Delta n$ and $\Delta\phi$ fluctuations unfavorable.  On the other hand, reduction of $\Delta n$ and $\Delta\phi$ is possible but bound by the important relation~\cite{sqzparam} 
\begin{equation}
 \Delta n\,\Delta\phi \gtrsim \frac 12\,,
\end{equation}
stating that we can in principle decrease one of the variances, however, at the cost of increasing the other one.  For a binomial state we have $\Delta n = \sqrt{N}/2$ and $\Delta\phi=1/\sqrt N$, leading to standard quantum shot noise~\cite{ober2}.  In contrast, for states with smaller $\Delta n$ or $\Delta\phi$ values, the so-called \textit{squeezed states}, we can achieve a measurement precision below standard quantum shot noise~\cite{ober1}.

In order to quantify how much a state is squeezed several factors have been used in the literature.  The so-called number squeezing and phase squeezing factors $\xi_n =\Delta n / (\sqrt{N}/2)$ and  $\xi_\phi = \Delta \phi / (1/\sqrt{N})$, respectively, provide information about how much a given state is squeezed in comparison to a binomial one.  Both factors equal one for a binomial state.  However, these factors completely neglect the coherence of the split condensate, a quantity of paramount importance for interferometry.  Coherence is additionally considered by the factor~\cite{sqzparam,pitaevskii:01}
\begin{equation}
  \alpha=\frac 2N\sqrt{\left<J_x \right>^2+\left< J_y \right>^2}=\left<\cos\phi\right>\,,
\end{equation}
where we have used the fact that $\left<\sin\phi\right>=0$ at equilibrium to arrive at the last expression on the right hand side.  It is now convenient to introduce the so-called coherent spin squeezing factor~\cite{sqzparam}
\begin{equation}
  \xi_S=\frac{\Delta n}{(\sqrt N/2)\alpha}=\frac{\xi_n}\alpha\,,
\end{equation}
which is a direct measure of the useful number squeezing in the context of interferometry.  In the following we will refer to states with low squeezing factors as ``highly squeezed states''.

Squeezed states cannot only be used for measurements with precisions beyond the standard quantum limit~\cite{ober1}, but also have other interesting properties.  For instance, number squeezed states are very robust against dephasing effects~\cite{jo:07}.  It is therefore important to find ways of producing number or phase squeezed states, ideally on short time scales.  A possible route towards number squeezing is to simply increase the distance between the two wells quasi-adiabatically~\cite{javanainen:99}: this reduces tunneling and, in turn, $\Delta n$, since the number squeezing in the groundstate of Eq.~\eqref{ham} increases with decreasing $\Omega$.  Additionally $\Delta \phi$ increases.  However, this process is relatively slow.  

In this paper we will explore a different approach towards highly phase or number squeezed states on short time scales, which relies on parametric amplification through a periodic modulation of the tunnel coupling with twice the resonance frequency.  In the following we briefly recall the mechanism underlying parametric amplification.  We start with the Hamiltonian of Eq.~(\ref{ham}) and rewrite it using the particle imbalance $n$ and the relative phase $\phi$~\cite{phaserep},
\begin{equation}
 H =  - \Omega\, \cos\phi+2\kappa\, n^2\,.
\label{ham2}
\end{equation}
In the coupled regime the relative phase is assumed to be very small, so we can approximate $\cos \phi\approx 1-\phi^2/2$.  Apart from an irrelevant constant energy shift, this expansion leads to
\begin{equation}
 H =  \frac\Omega2\ \phi^2 +2\kappa\, n^2\,.
\label{ham3}
\end{equation}
From the commutation relations of the spin operators one observes that $\phi$ and $n$ are canonically conjugate quantities, obeying $[\phi,n]=i$ \cite{phaserep}.  The Hamiltonian of Eq.~(\ref{ham3}) thus has exactly the same form as the Hamiltonian of a harmonic oscillator, with phase and number playing the role of momentum and position, and the ``mass'' of the oscillator given by $1/\Omega$.  Starting with a small amplitude at time zero, parametric amplification for the harmonic oscillator occurs for a time-dependent $\Omega$ which is modulated with twice the Josephson frequency $\omega_J = 2\sqrt{\kappa \Omega}$.  This leads to an exponential increase of the oscillator's amplitude, in case of Eq.~\eqref{ham3} the density $n$.  We emphasize that parametric amplification is also possible for higher $\phi$ values without performing a Taylor expansion of $\cos\phi$, as discussed in more detail in Ref.~\cite{smerzi97}.

Since for a split BEC we can modulate the tunneling parameter $\Omega$ by changing the distance between the wells, we can use parametric amplification in order to increase squeezing.  We first start with a slightly number squeezed groundstate of a split but still tunnel-coupled BEC in a double well.  In a next step, we modulate $\Omega$ with twice the Josephson frequency to get parametric squeezing amplification.  In contrast to the above example of the harmonic oscillator, the $\Omega$ modulation leads to an amplification of the fluctuations rather than the mean values.  

\section{Parametric squeezing amplification of a BEC}\label{sec:amplification}

In this section we show how to achieve parametric amplification for a BEC in a double well with realistic experimental parameters, in order to achieve high number or phase squeezing.  To describe the BEC correctly, a simple two mode model with static orbitals is not sufficient, as will be discussed below, and one has to resort to a description scheme that accounts for both the atom number and wavefunction dynamics.  In this work we employ the multi-configurational time-dependent Hartree method for bosons (MCTDHB)~\cite{mctdhb} using our recently developed Matlab toolbox OCTBEC~\cite{OCTBEC}.

\subsection{Simulation details}

\begin{figure}
\includegraphics[width=0.9\columnwidth]{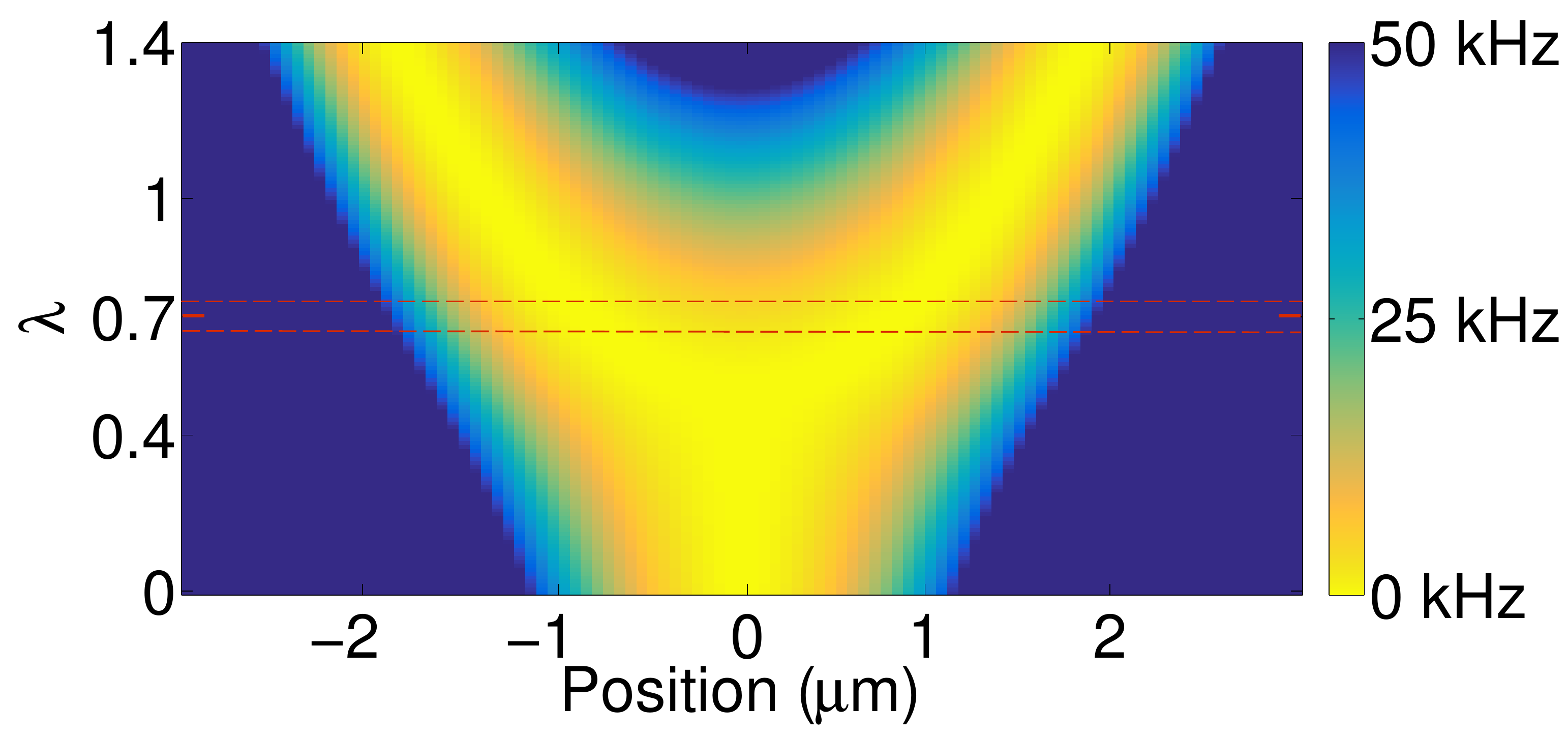}
\caption{(Color online) Lesanovsky-type potential~\cite{lesanovsky}, as used in our simulations, which allows to change from a single to a double well by modifying the control parameter $\lambda$ associated with rf fields of an atom chip.  The dashed lines indicate the 5\% modulations used in our simulations.  $\lambda$ primarily controls the distance between the two wells, but additionally also modifies the barrier height.}\label{les}
\end{figure}

\begin{figure}[b]
 \includegraphics[width=0.9\columnwidth]{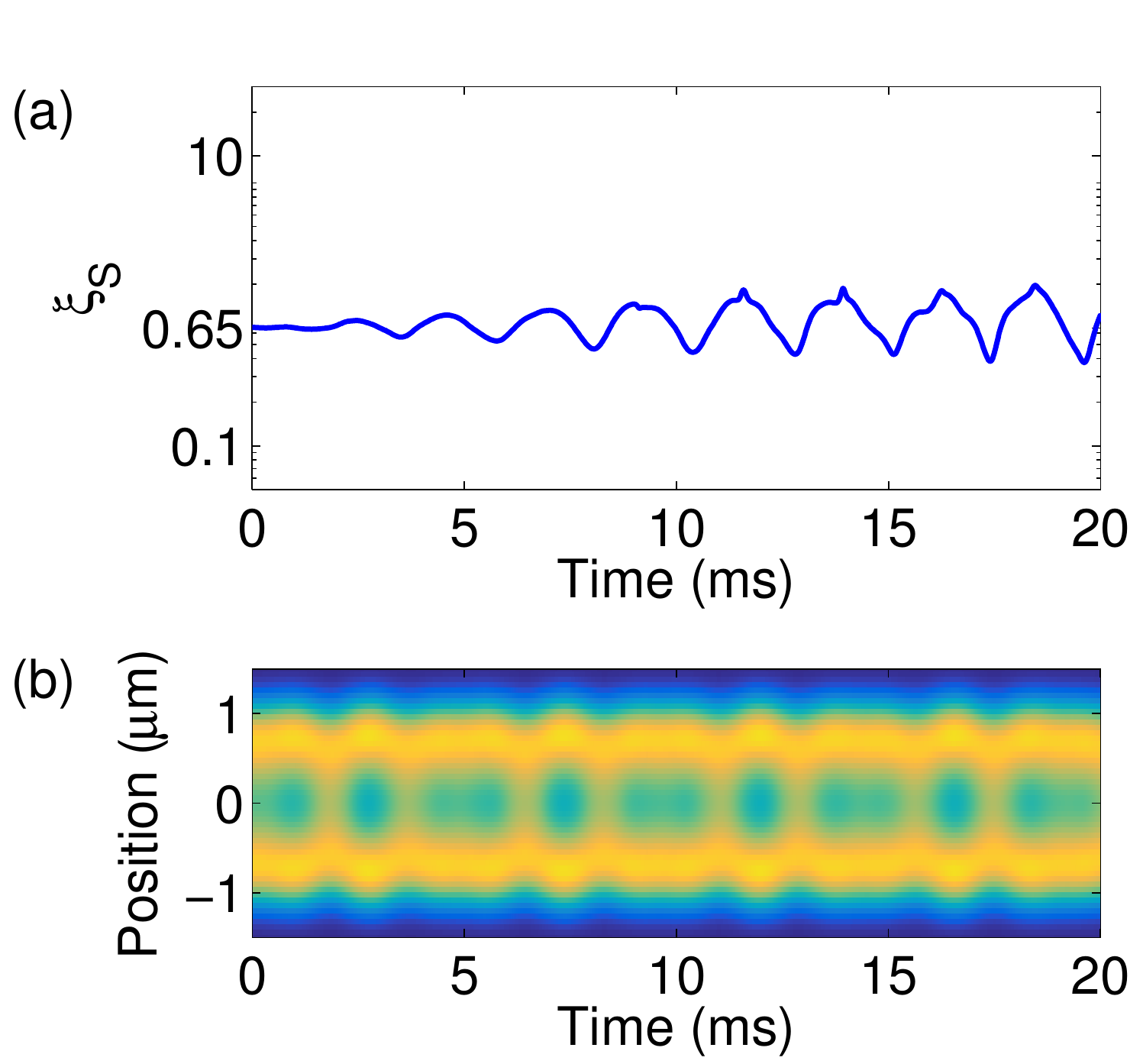}
 \caption{(Color online) Parametric amplification with an amplitude of the control parameter $\lambda$ of 1\%: (a) Coherent spin squeezing factor $\xi_S$ and (b) BEC density.  Same color bar as in Fig.~2.}
\label{pa1}
\end{figure}

\begin{figure}[b]
 \includegraphics[width=0.9\columnwidth]{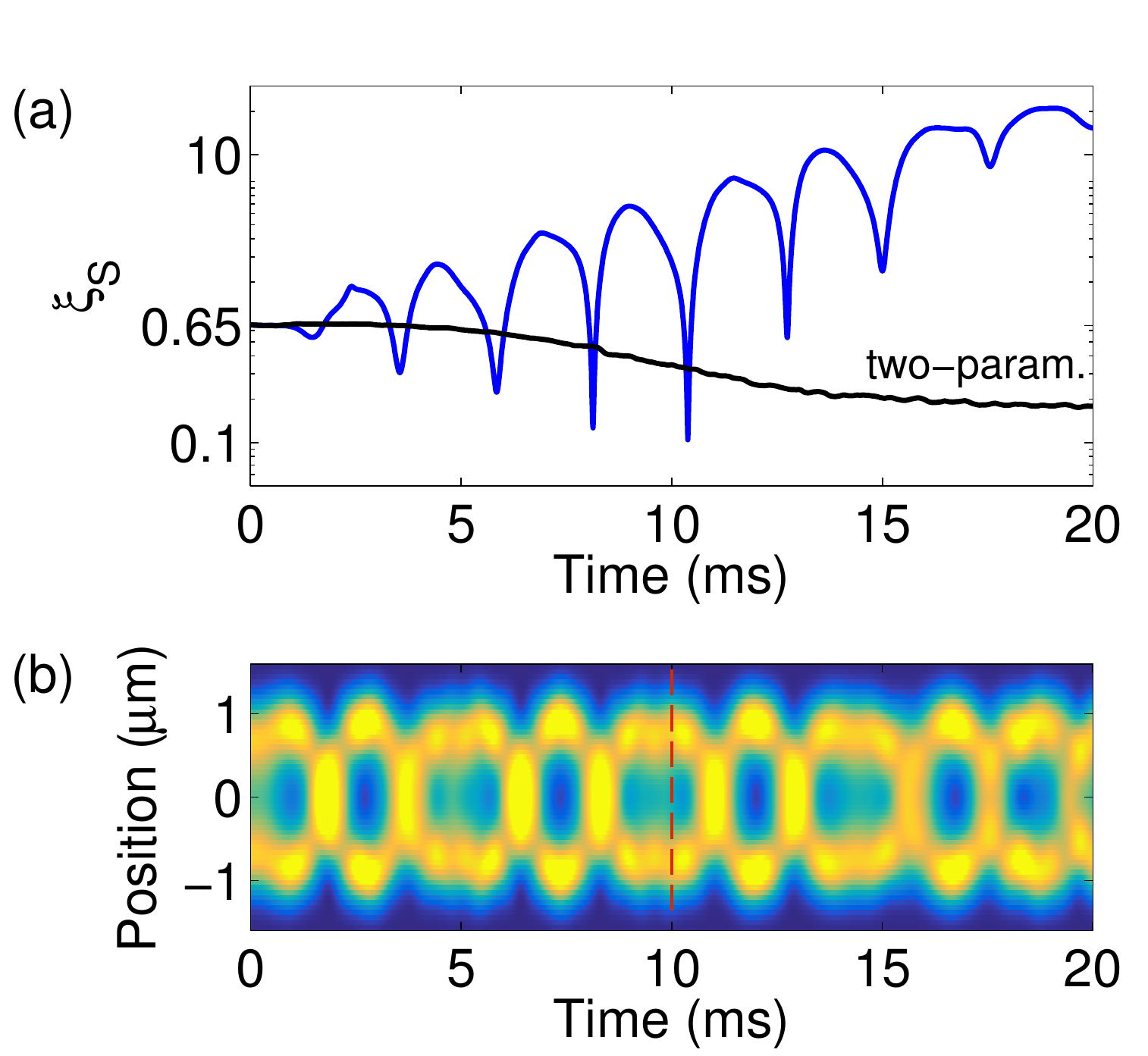}
\caption{(Color online) Same as Fig.~\ref{pa1} but for an amplitude of 5\%.  The black line reports for comparison results for a two parameter optimization~\cite{law:01}. }
\label{pa2}
\end{figure}

We simulate parametric amplification of a BEC consisting of 1000 atoms in a double well potential within MCTDHB(2)~\cite{mctdhb}, which expands the BEC wavefunction in two orbitals. The trap is a  Lesanovsky type potential~\cite{lesanovsky}, giving rise to elongated, cigar shaped condensates.  In this potential, the relevant splitting and amplification dynamics occurs in the radial direction, which allows us to introduce a 1d description scheme.  The Lesanovsky potential has a single parameter $\lambda$, associated with the amplitude of a radio frequency field, that can turn the trap from a single into a double well, as depicted in Fig.~\ref{les}.

\subsection{Parametric amplification}
\label{sec:pa}

In our simulations, we start with a BEC corresponding to the split ground state of the double well (we use $\lambda=0.7$).  It has been demonstrated experimentally that this state can be realized through adiabatic splitting of an elongated single atom trap, without generating significant heating or a noticable thermal fraction~\cite{integrated}.  Finite temperature effects might lead to a slight broadening of $\Delta\phi$ but will not significantly influence the spin squeezing factor $\xi_S$~\cite{ober2}.  In experiment, this might reduce the time that is avalaible for the parametric amplification.

In the split ground state the spin squeezing factor is initially already smaller than one ($\xi_S=0.65$ for $\lambda=0.7$).  Starting at time zero, the distance between the double well is modulated with twice the Josephson frequency $\omega_J/(2\pi) = 220$~Hz, giving rise to a parametric amplification of squeezing.  The squeezing value mainly depends on the amplitude of the modulation.  Figs.~\ref{pa1} and \ref{pa2} show the spin squeezing factor $\xi_S$ and the atomic density during parametric amplification for $\lambda$-modulations with amplitudes of 1\% and 5\%, respectively.  The density oscillates periodically for an amplitude of 1\%, while strong excitation and non-periodic features can be observed for 5\%. 

\begin{figure}
\centerline{\includegraphics[width=0.6\columnwidth]{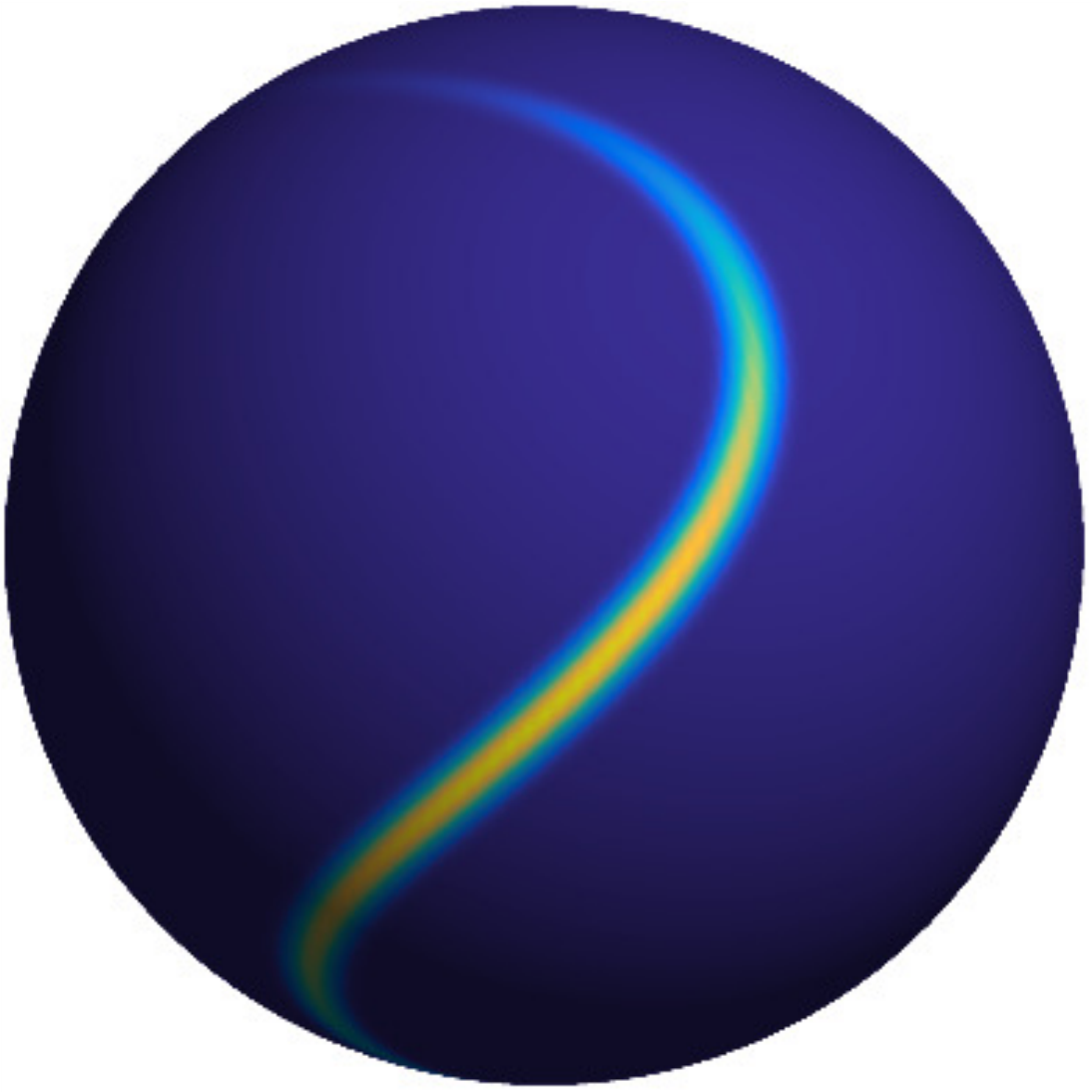}}
 \caption{(Color online) Snap shot of state that suffers dephasing, as obtained from the simulation with a 5\% modulation at time $t=12$ ms (see Fig.~\ref{pa2}).  Through parametric amplification wavefunction components with all atoms in the left or right well become populated, leading to a time evolution where the distribution ``curls'' around the Bloch sphere and squeezing is diminished.  See Fig.~2 for color bar.}
\label{curl}
\end{figure}

Parametric amplification with an amplitude of 1\% only produces a squeezing factor of $\xi_S \approx 0.4$ $(\xi_S^2 \approx -8$~dB), while the modulation with 5\% leads to a much better squeezing of $\xi_S \approx 0.1$ ($\xi_S^2 \approx -20 $~dB).  However, squeezing becomes worse after roughly 10 ms.  We attribute this degradation to dephasing effects: as depicted in Fig.~\ref{curl}, at later times the number distribution becomes curled around the $x$-axis of the Bloch sphere, indicating the partial occupation of states where all atoms reside in the left or right well, leading to complicated ensuing number dynamics with a net effect reminiscent of dephasing.  

A key requirement for interferometry on atom chips is a reliable and fast generation of squeezed states.  To boost parametric amplification on short time scales, one has to use sufficiently high tunnel coupling modulations, which, in turn, lead to excitations of the condensate wave function.  In this context, the consideration of both the atom number and wavefunction dynamics becomes mandatory in a simulation approach, thus calling for realistic many-body simulation approaches such as the MCTDHB framework of this work.  Additionally, following the parametric amplification one has to modify the trap potential in such a way that the orbitals are brought to a halt.  This step will be discussed in the next section.  The main advantage of parametric squeezing amplification compared to other routes towards number squeezing~\cite{julian09,grond09} is that the whole parametric amplification process can be implemented experimentally very easily, and only the final trapping stage requires some fine-tuning of the atom chip potentials.  For comparison, the black line in Fig.~\ref{pa2} reports results for a two parameter optimization~\cite{law:01}, whose final $\xi_S$ value is also comparable to genuine OCT protocols for the optimization of number squeezing~\cite{grond09,julian09}.  Note that in comparison to these optimized protocols our simple parametric amplification scheme already leads to higher squeezing.

\section{Condensate trapping}\label{sec:trap}

\begin{figure*}
\includegraphics[width=2\columnwidth]{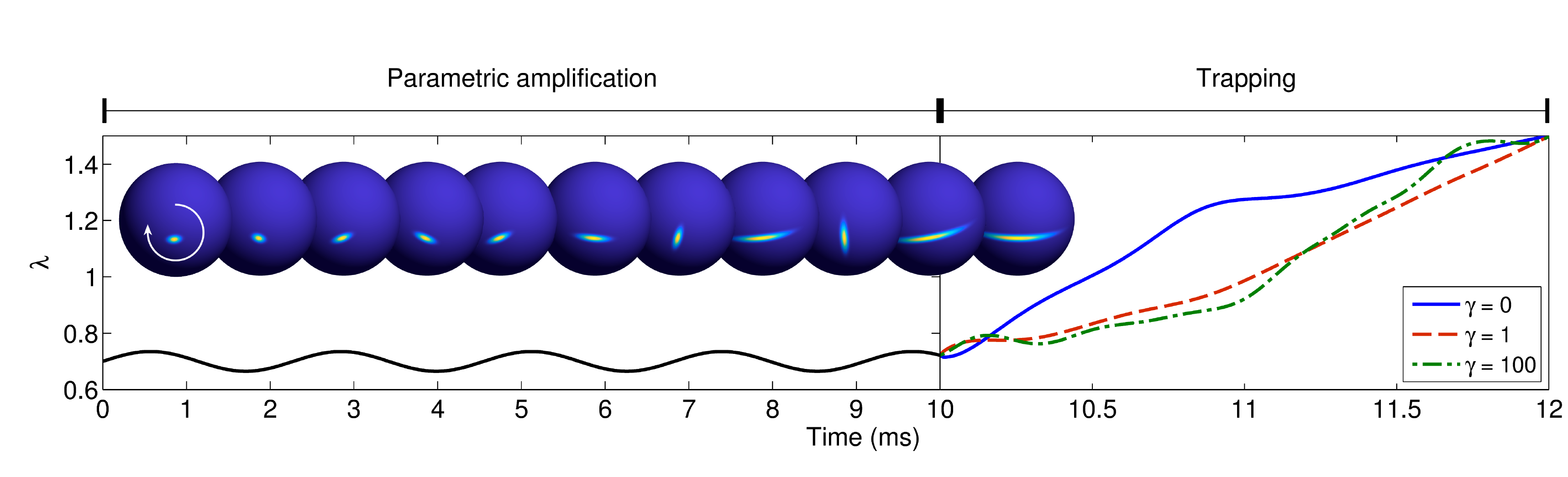}
\caption{(Color online) Parametric squeezing amplification (for times below 10 ms) and trapping (for times later than 10 ms, note different time axes).  We show the time evolution of the control parameter $\lambda$ for trapping ramps obtained for different weighting factors $\gamma$.  The inset shows Bloch sphere representations of states which rotate clockwise around the $x$-axis and become continuously squeezed.  See Fig.~2 for color bar.}
\label{trap1}
\end{figure*}

\subsection{Optimal control theory}
\label{sec:oct}

To make parametric amplification useful in the context of squeezing generation, we should be able to trap the BEC after amplification in a highly number squeezed state and separate the two wells far enough to inhibit interwell tunneling.  We will refer to the stage where the condensate is brought to a halt as `` trapping'', not to be confused with the atom trapping in order to produce a BEC on the atom chip.  Trapping is shown in Fig.~\ref{trap1} (for details see discussion below) and is accomplished within the framework of OCT.

The general goal of OCT is to solve the following inverse problem:  suppose that the state of a system $\Psi_0$ is known at some initial time $t_0$, and we are seeking for a desired state $\Psi_d$ at some final time $T$.  In order to bring the system from $\Psi_0$ to $\Psi_d$, we can tune some external control paramaters, such as the $\lambda$-parameter for the Lesanovsky potential.  In general, the time dependence of the control parameter that brings the system from the initial to the desired state is unknown.  OCT allows to find an optimal control in an iterative process.  Many variants of OCT implementations exist, such as CRAB~\cite{crab}, Krotov's method~\cite{krotov,sklarztannor}, or a gradient ascent pulse engineering (GRAPE) scheme~\cite{grond09,grapekrotov}.  In this work we employ the GRAPE algorithm implemented in the OCTBEC toolbox~\cite{OCTBEC}.

The OCT ingredients are the initial state of our system $\Psi_0$, a dynamic equation for the system's time evolution (here MCTDHB), and a cost function that rates the success for a given control field $\lambda(t)$.  For $\Psi_0$ we use the system's state after an initial parametric amplification stage.  As for the terminal cost, we are seeking for highly number squeezed states and for condensates at rest.  This can be accomplished through a cost function of the form
\begin{equation}
\mathcal{J}_T = \langle \Psi |J_z^2  |\Psi \rangle+ \frac{\gamma}{N} \langle \Psi | H | \Psi \rangle \,,
\label{costfun}
\end{equation}
which consists of two parts:  the first one favors strongly number-squeezed states, the second one minimal energy and thus a condensate at rest. $\gamma$ is a parameter that weights the relative importance for these two optimization goals. 

A slight complication arises for the squeezing term in Eq.~\eqref{costfun}, as $J_z$ is defined in the left-right basis whereas the natural MCTDHB basis is a gerade-ungerade basis~\cite{grond09}.  To switch between the two bases, we use
\begin{align}
 \phi_l = \frac{1}{\sqrt{2}} (\phi_g + \tilde f \phi_u)\phantom{\,,}\\
 \phi_r = \frac{1}{\sqrt{2}} (\phi_g - \tilde f \phi_u)\,,
\end{align}
where $\tilde f=f/|f|$ is the relative phase between the orbitals, which is obtained from the wave function overlap for $x>0$ ($\theta$ denotes the Heaviside step function)
\begin{align}
  f=\int \theta(x)~\phi_g^\ast(x)\phi_u(x)\, dx\,.
\label{phasef}
\end{align}
The constraint that the BEC dynamics is governed by the MCTDHB equations of motion is included within a  Lagrangian framework, and the full Lagrangian contains an additional cost term that favors control fields where the control parameter changes smoothly~\cite{OCTBEC},
\begin{equation}
\mathcal{L} =\mathcal{J}_T+\frac{\nu}{2}\int_0^T [\dot{\lambda}]^2 \,dt + \text{Re} \langle \tilde{\Psi}, i \dot{\Psi} - F(\Psi,\lambda)\rangle \,.
\label{lagrange}
\end{equation}
Here $F$ is a short-hand notation for the equations of motion, $\nu $ is a weighting parameter, and $\langle a,b \rangle$ is a short-hand notation for $\int_0^T dt\int dx\, a^\ast(x,t)b(x,t)$.

We next derive from this Lagrangian the optimality system that is needed for OCT.  With exception of the cost function, the pertinent equations for the MCTDHB approach can be found in Ref.~\cite{OCTBEC}, and we thus only comment on the functional derivatives of the terminal cost function $\mathcal{J}_T$.  Because of the relative phase $\tilde f$, see Eq.~\eqref{phasef}, appearing in the operator ${J}_z$, these derivatives are somewhat involved.  After some calculations, which are briefly sketched in appendix A, we arrive at
\begin{subequations}\label{eq:optimality}
\begin{align}
\frac{\partial \mathcal{J}_T}{\partial C^\ast}&=  J_z^2|C\rangle+\frac{\gamma}{N}~H|C\rangle \\
\frac{\partial \mathcal{J}_T}{\partial \phi_g^\ast}&= \langle C| J_z \frac{\partial J_z}{\partial \phi_g^\ast}+\frac{\partial J_z}{\partial \phi_g^\ast} J_z|C\rangle \\
\frac{\partial \mathcal{J}_T}{\partial \phi_u^\ast}&= \langle C| J_z \frac{\partial J_z}{\partial \phi_u^\ast}+\frac{\partial J_z}{\partial \phi_u^\ast} J_z|C\rangle\,,
\end{align}
\end{subequations}
with
\begin{subequations}
\begin{align}
\frac{\partial J_z}{\partial \phi_g^\ast}&= \frac{ \theta(x) \phi_u}{4} \left( \frac{a_g^\dagger a_u}{|f|}-a_u^\dagger a_g \frac{(f^\ast)^2}{|f|^3}\right )\\
\frac{\partial J_z}{\partial \phi_u^\ast}&= \frac{\theta(x) \phi_g}{4} \left( \frac{a_u^\dagger a_g}{|f|}-a_g^\dagger a_u \frac{(f)^2}{|f|^3}\right )\,.
\end{align}
\end{subequations}

For the optimizations we employ the Matlab toolbox OCTBEC~\cite{OCTBEC}.  See also Refs.~\cite{stateinversion,grapekrotov} for a detailed description of our OCT implementation.

\subsection{Trapping}
\label{sec:results}

In our OCT simulations we first perform a parametric amplification with an amplitude of 5\% and $t_0=10$ ms,\footnote{The success for optimizing squeezing and wavefunction trapping depends on the initial and terminal times $t_0$ and $T$, respectively.  $t_0=10$ ms was obtained from a linesearch, where we used a linear $\lambda$ ramp for trapping in order to find the ``best'' initial time in the interval $t_0\in[9,11]$ ms.  Also the length of the trapping sequence (here 2 ms) was optimized through a similar linesearch.} as shown in Figs.~\ref{pa2} and \ref{trap1}.  The system's state at this terminal time is then used for $\Psi_0$ in our OCT algorithm.  For the initial guess of the splitting and trapping ramp we use a linear ramp for $\lambda$ and a time interval of $T-t_0=2$ ms.  The initial guess was then optimized with the scheme described in Sec.~\ref{sec:oct} and for different weighting parameters $\gamma$. 

\begin{figure}
\includegraphics[width=\columnwidth]{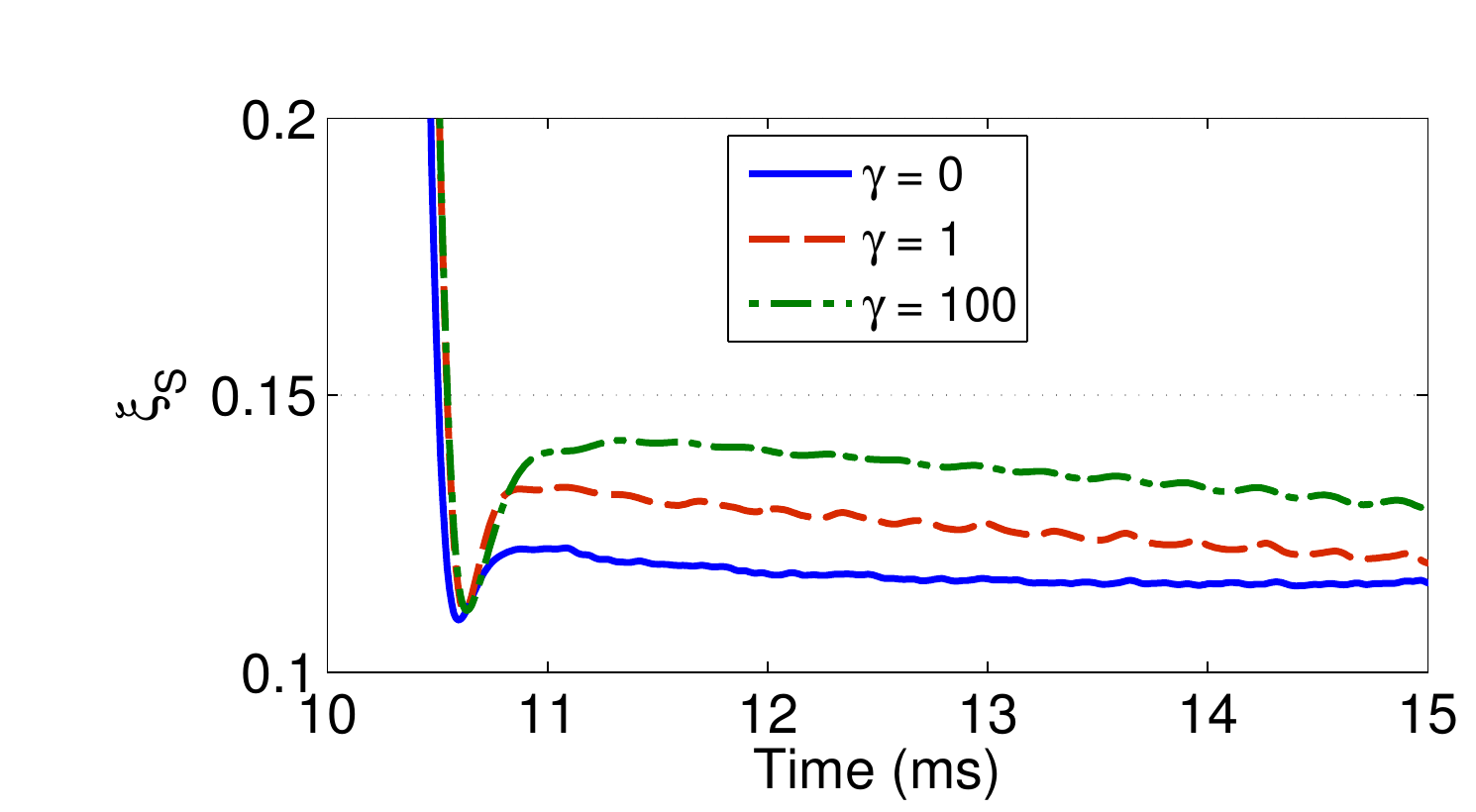}
\caption{(Color online) Coherent spin squeezing during and after trapping, for the ramps shown in Fig.~\ref{trap1}.  Smaller $\gamma$ values lead to better squeezing.}
\label{trap2}
\end{figure}

\begin{figure}
\includegraphics[width=\columnwidth]{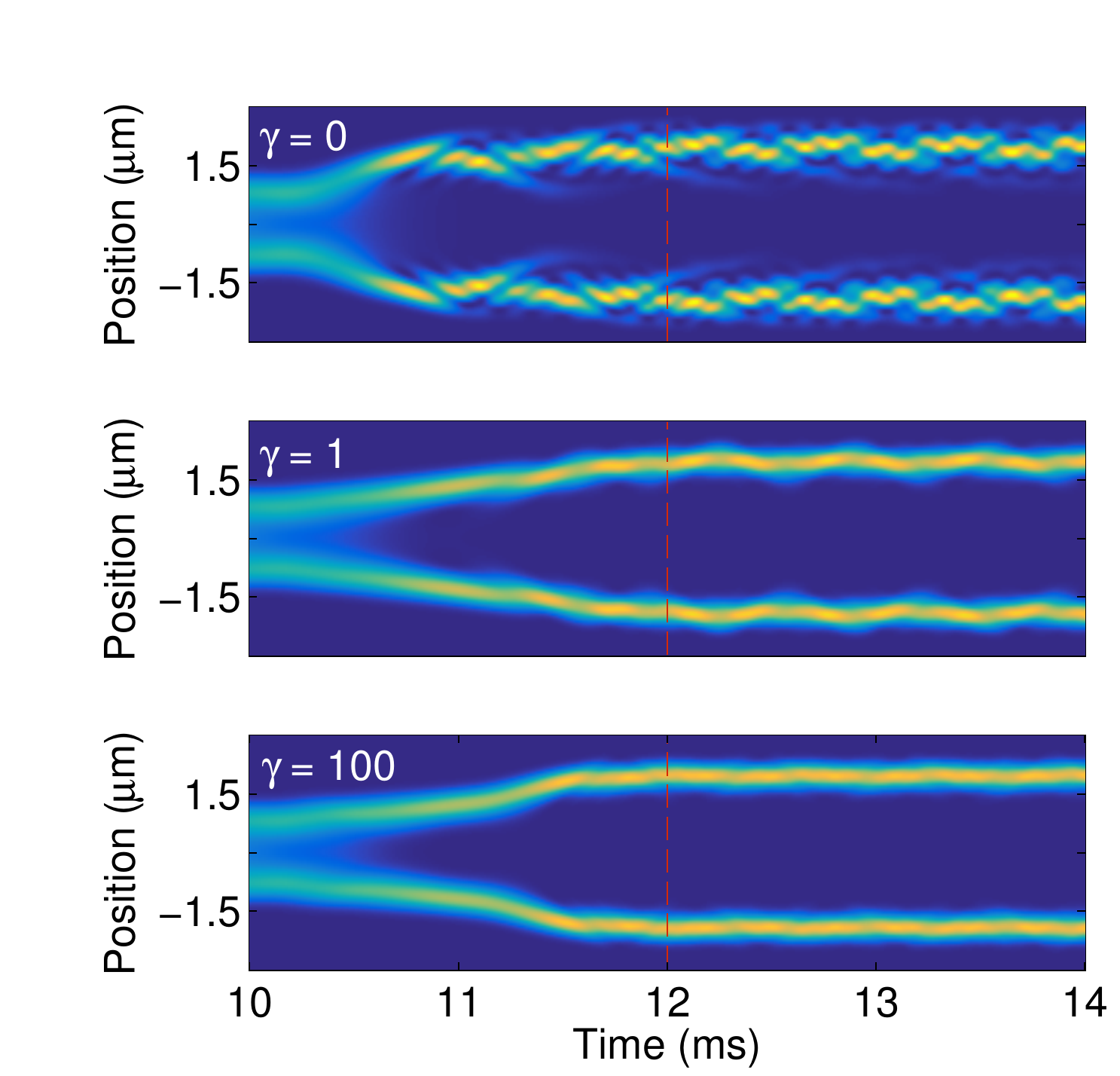}
\caption{(Color online) BEC density maps during and after trapping (dashed lines indicate end of trapping stage, see Fig.~2 for color bar) for the ramps shown in Fig.~\ref{trap1}. Larger $\gamma$ values lead to a less excited BEC.}
\label{trap3}
\end{figure}

Fig.~\ref{trap1} shows the resulting ramps for $\gamma$-factors of 0, 1, and 100.  For the additional cost penalization term in Eq.~\eqref{lagrange} we use a small value of $\nu=10^{-6}$ such that the control selection is only governed by $\mathcal{J}_T$ of Eq.~\eqref{costfun}.  The difference between these ramps is  attributed to the impact of the $\gamma$-factor that weights between the different optimization objectives of squeezing and trapping.  Fig.~\ref{trap2} depicts the resulting spin squeezing factors $\xi_S$ for the ramps shown in Fig.~\ref{trap1}, and Fig.~\ref{trap3} the corresponding atom densities.

All three ramps produce squeezing values lower than 0.13, corresponding to $\xi_S^2 \approx -18$~dB.  This is roughly 10~dB above the Heisenberg limit of $-28$ dB.  As expected, the squeezing values are better for optimizations with smaller $\gamma$ values, although the influence is not overly large.  From the density maps shown in Fig.~\ref{trap3} we infer that the ramp with $\gamma = 0$ leads to an excited BEC, the ramp with $\gamma = 1$ produces an only weakly excited BEC, and the ramp with $\gamma = 100$ results in a BEC that is almost at rest.

\section{Summary}\label{sec:summary}

We have discussed a parametric amplification scheme for creating and trapping a BEC in a highly squeezed state, with a squeezing value of $\xi_S^2 \approx -18$~dB.  Squeezing amplification is achieved in a split BEC through modulation of the tunnel coupling with twice the Josephson frequency.  To achieve high squeezing on short time scales, one has to use sufficiently large modulation amplitudes, which, in turn, lead to condensate oscillations.  These oscillation can be brought to halt through a splitting ramp optimized within the OCT framework.  Compared to other protocols for number squeezing~\cite{julian09,grond09}, the method presented here needs OCT only for the final trapping stage of the squeezed state.

\section*{Acknowledgements}

This work has been supported in part by the Austrian science fund FWF under project P24248 and by NAWI Graz.

\begin{appendix}

\section{}

We describe the BEC dynamics within the framework of MCTDHB(2)~\cite{mctdhb}.  In this method, the BEC wave function is expanded into a set of time dependent orbitals, which, for a spatially symmetric problem, can be classified according to their parity as gerade and ungerade, i.e., $\phi_g(x)$ and $\phi_u(x)$.  In order to find the optimality system given in section~\ref{sec:oct} we have to calculate all the derivatives of the cost function of Eq.~\eqref{costfun}, namely $\frac{\partial \mathcal{J}_T}{\partial C^\ast}$, $\frac{\partial \mathcal{J}_T}{\partial \phi_g^\ast}$, and $\frac{\partial \mathcal{J}_T}{\partial \phi_u^\ast}$.  The difficulty here is that the operator $J_z$ depends explicitly on the orbitals,
\begin{equation}
 J_z= \frac{1}{2} \left( \tilde f a_g^\dagger a_u + \tilde f^\ast a_u^\dagger a_g\right),
\end{equation}
namely through the factor $\tilde f$ that depends on $\phi_g$ and $\phi_u$, see Eq.~\eqref{phasef}.  Performing the functional derivative $\frac{\partial \mathcal{J}_T}{\partial C^\ast}$ is straightforward, and we arrive at 
\begin{equation}
 \frac{\partial}{\partial C^\ast} \left(\langle C|J_z^2|C\rangle+\frac{\gamma}{N}~\langle C|H|C\rangle \right) =J_z^2|C\rangle+\frac{\gamma}{N}~H|C\rangle \,.
\end{equation}
For $\frac{\partial \mathcal{J}_T}{\partial \phi_g^\ast}$ the second term of the cost function vanishes, since there is no dependence on the orbitals.  We start by using the chain rule
\begin{align}
\frac{\partial \mathcal{J}_T}{\partial \phi_g^\ast}=\frac{\partial \mathcal{J}_T}{\partial J_z} \frac{\partial J_z}{\partial \phi_g^\ast}=\langle C| J_z \frac{\partial J_z}{\partial \phi_g^\ast}+\frac{\partial J_z}{\partial \phi_g^\ast} J_z|C\rangle \,.
\end{align}
To calculate $\frac{\partial J_z}{\partial \phi_g^\ast}$ we first use
\begin{displaymath}
\frac{\partial f}{\partial \phi_g^\ast} = \phantom{-}\frac{1}{2|{f}|}\theta(x) \phi_u\,,\quad
\frac{\partial f^\ast}{\partial \phi_g^\ast} = - \frac{({f}^\ast)^2}{2|{f}|^3}\theta(x) \phi_u\,,
\end{displaymath}
and arrive at 
\begin{align}
\frac{\partial J_z}{\partial \phi_g}&= \frac{\theta(x) \phi_u}{4} \left( \frac{a_g^\dagger a_u}{|f|}-a_u^\dagger a_g \frac{(f^\ast)^2}{|f|^3}\right )\,.
\end{align}
The calculation of $\frac{\partial J_z}{\partial \phi_u^\ast}$ is very similar and we find
\begin{align}
\frac{\partial J_z}{\partial \phi_u}&= \frac{\theta(x) \phi_g}{4} \left( \frac{a_u^\dagger a_g}{|f|}-a_g^\dagger a_u \frac{(f)^2}{|f|^3}\right )\,.
\end{align}
This leads us to our final result of Eq.~\eqref{eq:optimality}.

\end{appendix}


\end{document}